\documentstyle[12pt,amssymb]{article}
\newtheorem{theorem}{Theorem}

\begin{document}

\begin{titlepage}
\begin{center}
\begin{Large}
{\bf Quantum Mechanics on the Hypercube}
\end{Large}

\vskip3truecm

E. G. Floratos$^{(a)}$\footnote{E-mail:
manolis@timaios.nuclear.demokritos.gr} and
S. Nicolis$^{(b)}$\footnote{E-mail:
nicolis@celfi.phys.univ-tours.fr. Partially supported by the GDR 682
``Structures Non-Perturbatives en Th\'eories de Cordes et Th\'eories
de Champs''.}

\vskip1truecm
$^{(a)}$ {\sl Institute of Nuclear Physics, NRCPS ``Demokritos''\\
15310 Aghia Paraskevi, Athens, Greece}

$^{(b)}$  {\sl CNRS--Laboratoire de Math\'ematiques et Physique
Th\'eorique (UPRESA 6083)\\
Universit\'e de Tours, Parc Grandmont, 37200 Tours, France}

\vskip5truecm

\begin{abstract}
We construct  quantum evolution operators on the space of states, that is
represented by the vertices of the $n$-dimensional unit hypercube. 
They realize the metaplectic representation of the modular group
$SL(2,{\Bbb Z}_{2^n})$. 
By construction this representation acts in a natural way on the
coordinates of the non-commutative 2-torus, ${\Bbb T}^2$ and thus is 
 relevant for noncommutative field theories as well as theories of
quantum space-time.

\end{abstract}
\end{center}

\end{titlepage}
Recent progress in M-theory indicates that spacetime itself becomes 
noncommutative at scales where D-branes play an important
role~\cite{matrixmodel,schwarz}.
This noncommutativity comes about in a rather natural way because
D-branes are charged, gravitational solitons, moving in backgrounds
with magnetic flux. This is reminiscent of the  Landau problem 
\cite{connes}. 

On the other hand, quantum mechanics provides a prototype of
noncommutativity--but in phase space. In the Landau problem 
the noncommutativity of the two, real, space coordinates is brought
about by the magnetic flux.

In previous work~\cite{fqmdikamas,fastqmaps,su2}, taking advantage of the existence
of finite dimensional representations of the Heisenberg-Weyl group,
for specific values of Planck's constant, {\em viz.} $\hbar=2\pi/N$, 
we studied quantum spaces~\cite{madore}, i.e. whose coordinates are
(finite dimensional) matrices in this group. The object of this
exercise was to quantize, exactly, linear cellular automata, using the
metaplectic representation of $SL(2,{\Bbb Z}_N)$, where $N$ was any
odd integer. These quantum maps have been studied, in particular
within the context of quantum chaos~\cite{qchaos}, Rational Conformal
Field Theory~\cite{rcft} and quantum gravity~\cite{thooft}.

The case $N=2^n$ was not amenable to analysis using the tools thus far 
available, although it is of clear interest for quantum
computing~\cite{qualgor} and the state space has been widely used in
communication engineering. Indeed the principal
difficulty resides in resolving ambiguities due to the 
factors of $1/2$ that abound in the expressions of the metaplectic
representation. 

In this note 
we present a prescription that resolves these ambiguities and 
we construct the metaplectic representation of the
symplectic group (linear, canonical, transformations) for the discrete 
torus, ${\Bbb Z}_{2^n}\times {\Bbb Z}_{2^n}$~\cite{wolfram}

Our results are useful for studying field theories on noncommutative
spaces~\cite{seiberg}, whose coordinates are finite dimensional
matrices~\cite{ambjorn}, as well as fast quantum
algorithms~\cite{qualgor}.  

We begin by reviewing some key features of $SL(2,{\Bbb Z}_{2^n})$.

The classical evolution of a linear cellular automaton on the phase
space ${\Bbb Z}_{2^n}\times {\Bbb Z}_{2^n}$ is given by the discrete,
canonical, transformation
\begin{equation}
\left(\begin{array}{c} q_{n+1} \\ p_{n+1} \\ \end{array}\right)=
{\sf A}\cdot 
\left(\begin{array}{c} q_{n} \\ p_{n} \\ \end{array}\right)
\end{equation} 
where ${\sl A}\in SL(2,{\Bbb Z}_{2^n})=Sp(1,{\Bbb Z}_{2^n})$. 
The properties of this group can be analyzed using those of the
(abelian) lattice ${\Bbb Z}/{\Bbb Z}_{2^n}$. This lattice contains a
subrgoup, consisting of all odd integers mod $2^n$;it has $2^{n-1}$ 
elements. The main novelty, that distinguishes it from the case of $N$
odd is that the equation 
\begin{equation}
\label{rootsofunity}
x^2\equiv 1\,{\mathrm mod}\,2^n
\end{equation}
admits two new solutions
\begin{equation}
\label{newroots}
x_{\pm}=2^{n-1}\pm 1
\end{equation}
in addition to the old $\pm 1$, so there are {\em four} units. 

The group thus splits into two subgroups: the set consisting of the
two non-trivial units
and a cyclic subgroup of $2^{n-2}$ elements, whose primitive element is 3 or 5.

A subgroup of $SL(2,{\Bbb Z}_{2^n})$ of particular interest is the ``rotation 
group'', $SO(2,{\Bbb Z}_{2^n})$, comprising matrices of the form
\begin{equation}
\label{rotgroup}
{\sf A}=\left(\begin{array}{rr} a & b \\ -b & a\\ \end{array}\right)
\end{equation}
with $a^2+b^2\equiv 1\,{\mathrm mod}\,2^n$. 
Its structure is relevant for the study of the harmonic oscillator within 
the framework of Finite Quantum Mechanics and was studied in
Ref.~\cite{fqmdikamas}

We can find all elements of this group by solving
 the constraint, $a^2+b^2\equiv 1\,
{\mathrm mod}\,2^n$ in terms of an even number $t$
\begin{equation}
\label{sol1}
\begin{array}{c}
\displaystyle a=2t(t^2+1)^{-1}\,{\mathrm mod}\,2^n\\
\\
\displaystyle b=(t^2-1)(t^2+1)^{-1}\,{\mathrm mod}\,2^n\\
\end{array}
\end{equation}
This parametrization, together with that obtained by the exchanges 
$a\to b,\,b\to a$ and $a\to -b,\,b\to a$, gives us all the $2^{n+1}$
elements.

There is, in particular, a cyclic subgroup of dimension $2^{n-1}$,
generated by the element corresponding to the value $t=2$ for any
$n$\footnote{This primitive element is called the Balian-Itzykson oscillator.}.

After this introduction, 
we shall propose an explicit expression for the evolution operator,
$U({\sf A})$, ${\sf A}\in SL(2,{\Bbb Z}_{2^n})$ 
and we shall prove that it 
realizes (a) a group representation
and (b) a metaplectic representation. 

We define
\begin{equation}
\label{roots}
\begin{array}{c}
\displaystyle\omega_n=e^{2\pi{\mathrm i}/2^n}\\
\displaystyle\widehat{\omega}_n=e^{2\pi{\mathrm i}/2^{n+1}}\\
\end{array}
\end{equation}
and posit the following expression
\begin{equation}
\label{evolutionU}
U\left(\left[\begin{array}{cc} a & b \\ c & d\\ \end{array}\right]\right)_{k,l}=
\frac{c_n({\sf A})}{\sqrt{2^n}}\widehat{\omega}_n^{(ak^2-2kl+dl^2)/c}
\end{equation}
where 
$$
{\sf A}=
\left(\begin{array}{cc} a & b \\ c & d\\ \end{array}\right)\in 
SL(2,{\Bbb Z}_{2^n})
$$
The constant $c_n({\sf A})$ will be determined by the requirement that 
$U$ realize a group representation
$$
U({\sf A}\cdot {\sf B})=U({\sf A})\cdot U({\sf B})
$$
We note here that the expression for $U({\sf A})$ used for $N$ odd  
depends on $\omega_n$ and {\em not} on $\widehat{\omega}_n$, since 
$1/2$ existed mod $N$--whereas it is no longer defined. 

The proof of point (a) proceeds as follows:

We set ${\sf A}_{ij}=a_{ij}$, ${\sf B}_{ij}=b_{ij}$ and $({\sf A}\cdot {\sf B})_{ij}=c_{ij}$ and 
easily find 
\begin{equation}
\label{grouplaw}
\sum_{m=0}^{2^n-1}U({\sf A})_{k,m}U({\sf B})_{m,l}=\frac{c_n({\sf A}\cdot{\sf B})}{c_n({\sf A})c_n({\sf B})}\times
U({\sf A}\cdot {\sf B})_{k,l}\times
\frac{1}{\sqrt{2^n}}\sum_{m=0}^{2^n-1}
\widehat{\omega}_n^\phi
\end{equation}
where
$$
\phi=\frac{c_{21}}{a_{21}b_{21}}m^2
$$
The Gau\ss~sum,
\begin{equation}
\sigma_n(a)=\frac{1}{\sqrt{2^n}}\sum_{m=0}^{2^n-1}\omega_n^{am^2}
\end{equation}
takes the following values, for $N=2^n$~\cite{Lang}
\begin{itemize}
\item
$$\sigma_n(1)=1+{\mathrm i}$$
for $n>2$. 
\item
If $a$ is an odd integer, 
$$
\sigma_n(a)=(-2^n|a)\varepsilon(a)(1+{\mathrm i})
$$
where 
we have used  the symbol
$$
\varepsilon(a)=\left\{\begin{array}{c} 1\,\,\,\,a\equiv 1\,{\mathrm mod}\,4\\
                                 {\mathrm i}\,\,\,\,a\equiv 3\,{\mathrm mod}\,4\\

\end{array}
\right.
$$
\end{itemize}
and the Jacobi symbol\footnote{The notation $\left(\frac{a}{b}\right)$ is also
used.} 
$$
(a|b)=\left\{\begin{array}{r} 1\,\,\,\,a\equiv x^2\,{\mathrm mod}\,b\\
                             -1\,\,\,\,a\not\equiv x^2\,{\mathrm mod}\,b\\
\end{array}
\right.
$$
It is also possible to show that 
$$
\sum_{m=0}^{2^n-1}\widehat{\omega}_n^{m^2a}=\sqrt{2^{n-1}}\sigma_{n+1}(a)
$$
More properties of this sum may be found in Lang's text.

Choosing
the constant $c_n({\sf A})$ in eq.~(~\ref{evolutionU}) 
$$
c_n({\sf A})=-\frac{{\mathrm i}\sigma_{n+1}(c)}{\sqrt{2}}
$$
it follows that 
the evolution
operator 
\begin{equation}
\label{faithfulgrouprep}
U(A)_{k,l}=\frac{-{\mathrm i}\sigma_{n+1}(c)}{\sqrt{2^{n+1}}}
\widehat{\omega}_n^{(ak^2-2kl+dl^2)/c}
\end{equation}
realizes an $N=2^n$-dimensional representation, which provides 
the {\em exact} quantization of all linear cellular automata, ${\sf
A}\in SL(2,{\Bbb Z}_{2^n})$.

In the above derivation we assumed that the element $c={\sf A}_{21}$ is 
odd, in order that its inverse in the exponent be defined. 
When $c={\sf A}_{21}$ is even, then both $a={\sf A}_{11}$ and $d={\sf
A}_{22}$ must be odd (since $ad-bc\equiv\,1\,{\mathrm mod}\,2^n$). 
In this case we {\em define} $U({\sf A})$ as
\begin{equation}
\label{consistentdef1}
U({\sf A})=U({\sf A}\cdot {\sf S})\cdot
U^{-1}({\sf S})
\end{equation}
where 
$$
{\sf S}=\left(\begin{array}{rr} 0 & 1 \\ -1 & 0\\ \end{array}\right)
$$
This definition, for $c={\sf A}_{21}$ even, is chosen in order to
cover all the cases of the products of any two matrices, ${\sf A},{\sf
B}\in SL(2,{\Bbb Z}_{2^n})$ and guarantees that $U$ indeed defines a
representation. 

One nontrivial check of this is provided by the result
\begin{equation}
\label{newperiod}
U({\sf A})^{N/2}=U({\sf A}^{N/2})=U(I_{2\times 2})=I_{2^n\times 2^n}
\end{equation}
for ${\sf A}\in SO(2,{\Bbb Z}_{2^n})$,
which shows that the space of eigenstates splits into two copies,
indexed by a parity quantum number. This is to be contrasted to the
result obtained for $N=4k\pm 1$ prime, where $U({\sf
A})^{4k}=I_{N\times N}$. 
Eq.~(\ref{newperiod}) follows from the theorem:
\begin{theorem}
For $N=2^n$, $n>2$ and ${\sf A}\in SO(2,{\Bbb Z}_N)$ 
$$
{\sf A}^{N/2}\equiv I_{2\times 2}\,{\mathrm mod}\,N
$$
\end{theorem}
{\em Proof}: This is established by induction. 
Setting ${\sf a}_1=a,{\sf b}_1=b$, we find that 
\begin{equation}
\begin{array}{c}
\displaystyle {\sf a}_{k+1}={\sf a}_k a - {\sf b}_k b \\
 \displaystyle {\sf b}_{k+1}={\sf a}_k b + {\sf b}_k a \\
\end{array}
\end{equation}
For $n=3$, we find that 
\begin{equation}
{\sf a}_4=8a^4-8a^2+1\equiv 1\,{\mathrm mod}\,8
\end{equation}
which proves the first step. Assuming the theorem holds for some
$n>3$,  
we will show it holds for $n+1$, {\em viz.} 
\begin{equation}
{\sf a}_{2^{n-1}}\equiv 1\,{\mathrm mod}\,2^n\Rightarrow
{\sf a}_{2^n}\equiv 1\,{\mathrm mod}\,2^{n+1}
\end{equation}
Indeed, 
\begin{equation}
\begin{array}{c}
\displaystyle
{\sf a}_{2^n}={\sf a}_{2^{n-1}}^2-{\sf b}_{2^{n-1}}^2=2{\sf
a}_{2^{n-1}}^2-1 ; \\
\displaystyle
{\sf a}_{2^{n-1}}=1+2^n r\Rightarrow 2{\sf a}_{2^{n-1}}^2-1=
2^{n+1}(2^n r^2+2r)+1\equiv 1\,{\mathrm mod}\,2^{n+1}\\
\end{array}
\end{equation}
We note that ${\sf a}_{l}$, as a function of $a$,is a Chebyshev
polynomial of degree $l$. The theorem just proven is equivalent to 
 the amusing property that the
value of the order $l=2^n>4$ polynomial, for any, even, value of its
argument, is equal to $1\,{\mathrm mod}\,2^{n+1}$.

The last step of our construction is to prove 
that the evolution operator $U({\sf A})$,
constructed above, 
has the  {\em metaplectic} property, thus providing the exact
quantization of the action of linear, classical, maps ${\sf A}\in SL(2,{\Bbb
Z}_{2^n})$. 
We start by recalling  the definition of the 
generators of the Heisenberg--Weyl group 
\begin{equation}
\label{heisen-weyl}
J_{r,s}=\widehat{\omega}_n^{r\cdot s}P^r\cdot Q^s,\,\,\,\,r,s=0,\ldots,2^n-1
\end{equation}
where
$$
\begin{array}{c}
P_{k,l}=\delta_{k-1,l},\,\,\,\,k,l=0,\ldots,2^n-1\\
\\
Q_{k,l}=\omega_n^{k}\delta_{k,l},\,\,\,\,k,l=0,\ldots,2^n-1\\
\\
\end{array}
$$
The  metaplectic property, defined by the relation
\begin{equation}
\label{metaplectic}
U({\sf A})J_{r,s}U^{-1}({\sf A})=J_{(r,s){\sf A}}
\end{equation}
guarantees that $U({\sf A})$ quantizes the action of ${\sf A}$ on the
classical phase space. 

It  may be easily checked by a direct computation, that the operator,
given by eq.~(\ref{faithfulgrouprep}) indeed satisfies this definition. 

Furthermore, since the representation of the Heisenberg-Weyl group is
irreducible and the representation of $SL(2,{\Bbb Z}_{2^n})$
just constructed satisfies  the metaplectic property, 
we deduce that this representation is irreducible.  

This paper ends the cycle~\cite{fqmdikamas,fastqmaps}. 
 All the necessary 
tools, for dealing with the evolution operator of any quantum system 
within the framework of finite quantum mechanics, are now operative.
Indeed, for any value of $N$, using the Chinese remainder theorem, for
the prime factors $2^n\times N'$, where $N'$ is odd, it is possible to 
construct the metaplectic representation of $SL(2,{\Bbb
Z}_N)=SL(2,{\Bbb Z}_{2^n})\times SL(2,{\Bbb Z}_{N'})$ by tensoring the 
representations of the corresponding factor groups~\cite{fastqmaps}.   
They open several directions that merit further investigation:

In recent work on noncommutative gauge theory~\cite{schwarz,seiberg,ambjorn}
the coordinates satisfy 
non-trivial commutation relations, 
$[X_{\mu},X_{\nu}]=C_{\mu\nu}$. Our work indicates that it may also be
of interest to consider, instead, position operators that take values
in the group and not the algebra, that is impose 
$X_{\mu}X_{\nu}=\omega X_{\nu}X_{\mu}$, where $\omega$ is a root of
unity (Manin's quantum plane, cf.~\cite{madore}). 
In this, more general situation, one
might expect to recover at will classical/quantum dynamics and/or
(non)commutative  geometric effetcs, depending on how the scaling
limit is taken (i.e. what is held fixed as $N\to\infty$). 
A simple, but non-trivial, example is given by the Azbel-Hofstadter 
Hamiltonian~\cite{su2} that describes a charged particle moving on
the plane in the presence of a uniform transverse magnetic field and a 
periodic scalar potential, whose spectrum, for finite, rational
flux/plaquette, is generated by deformations of the $SU(2)$ algebra,
which, in the limit of infinitesimal flux, becomes a classical
$SU(2)$  algebra  for the quantum dynamics. 
Using elements of the Heisenberg-Weyl group as coordinates it would be
interesting to calculate, in perturbation theory, properties of
noncommutative Yang-Mills theories, because the finite dimension of
the Heisenberg-Weyl group provides, at the same time, an UV and IR cutoff~\cite{schwarz,seiberg}.

Another interesting application concerns 
quantum algorithms, which,  to date~\cite{qualgor},  are based mainly
 on the Fourier transform, which is related to 
the evolution operator of the harmonic oscillator.
We have considered arbitrary quantum maps, for which the
corresponding algorithms remain to be constructed. 

{\bf Acknowledgements:} We would like to acknowledge the warm
hospitality of the Laboratoire de Physique Th\'eorique de l'Ecole
Normale Sup\'erieure, where much of this work was carried out.

\end{document}